# Full dynamical solution for a spherical spin-glass model

L. F. Cugliandolo and D. S. Dean

*Service de Physique de L'Etat Condensé, Saclay, CEA*

*Orme des Merisiers, 91191, Gif–sur–Yvette Cedex, France.*

(February 17, 1995)

# Abstract

We present a detailed analysis for the Langevin dynamics of a spherical spin-glass model (the spherical Sherrington-Kirkpatrick model). All the spins in the system are coupled by pairs via a random interaction matrix taken from the Gaussian ensemble.

One finds that for a general initial configuration the system never reaches an equilibrium state and the theorems associated to 'equilibrium dynamics' are violated. Only very particular initial conditions drive the system to equilibrium.

The weak ergodicity breaking scenario is demonstrated for general 'non-equilibrium' initial conditions. Two-time quantities such as the autocorrelation function explicitly depend on both times. When the time difference is short compared to the smaller time one finds stationary dynamics with time translation invariance and the fluctuation-dissipation theorem satisfied. Instead, when the time difference is of the order of the smaller time one finds non-stationary dynamics with aging phenomena, the system *remembers* the time spent after the initial time (the quench below the critical temperature). Interestingly enough the short time-difference dynamics (FDT regime) for non-equilibrium initial conditions is identical to the relaxation within the equilibrium states obtained for the particular 'equilibrium' initial conditions. This points to a self-similariry of the energy landascape.

In addition we analyse the effect of temperature variations on the behaviour of the correlation function and energy density. In this way we are able to make some comparisons between experimentally observed results and exact calculations for the model. Somewhat surprisingly, this simple model captures some of the effects seen in laboratory spin-glasses.

# I. INTRODUCTION

Spin–glasses, like many other complex systems[1], are essentially out of equilibrium on experimental scales. Below the critical temperature experiments show 'aging effects', or the dependence of the response of the samples on their history since the temperature quench[2,3]. The simplest way to observe aging phenomena is through the zero-field cooling (ZFC) experiment in which the sample is cooled in zero field to a sub-critical temperature at time $t_o$; after a waiting time $t_w$ a small constant magnetic field is applied and subsequently the time-dependence of the magnetisation is recorded. Its 'mirror' counterpart consists of cooling the sample at the the quenching time $t_o = 0$ in the presence of a small constant magnetic field, keeping the sample in the field up to the waiting time $t_w$ and then measuring the decay of the thermoremanent magnetisation (TRM). In both cases, the 'older' the systems (the longer the waiting times) the slower their relaxations: the systems age[2,3].

Other, more sophisticated, experiments include variations of the temperature, mainly temperature jumps and temperature cyclings below the critical temperature $T_c$ during the waiting time. In the temperature jump experiments[4,6,7] the samples are kept at a constant temperature $T$ during a waiting interval $[0, t_j]$. At $t_j$ the temperature is changed to $T + \delta T$. The sample is still kept waiting at this constant temperature during the time interval $[t_j, t_w]$. Finally, at $t_w$ the magnetic field is switched on or off and the ZFC or TRM decays are measured, respectively, at the final temperature $T + \delta T$. In the simplest setting[4] $t_j = t_w$.

In the temperature cycling experiments, a temperature cycle is performed during the total waiting time; $T(t)$ is $T$ for times between $t_o$ and $t_{w1}$, $T + \delta T$ for times between $t_{w1}$ and $t_{w2}$ and again $T$ for all subsequent times. The magnetic field is switched at $t_w > t_{w2}$ according to the ZFC or TRM settings[5,6].

In all these experiments, the notion is to find an effective waiting time corresponding to a system that has undergone temperature variations. This effective waiting time can be estimated by comparing the decays of quantities such as TRM with the same system kept at constant measuring temperature. Whether to find the effective waiting time is possible depends on the precise protocol of the experiment[3].

The ZFC and TRM experiments[4,5] with temperature jumps at $t_j = t_w$ unequivocably show that waiting at a higher temperature ($\delta T < 0$) disfavours aging and makes the materials respond as younger ones while waiting at a lower temperature ($\delta T > 0$) favours aging and makes the materials respond as older ones[4] as to compared to a system kept at a constant measurement temperature $T + \delta T$. The response depends on the magnitude and duration of the heat pulses. The position of the maximum of the logarithmic derivative of the magnetisation decay is sometimes associated to an effective age of the system. After temperature jumps one observes a displacement of this maximum towards smaller (bigger) times[4] when $\delta T > 0$ ($\delta T < 0$).

*Grosso modo* this is also observed by comparing the full TRM decays with and without jumps[6,7]. However, a more detailed analysis of the TRM decay curves after temperature variations shows more subtle effects and provokes controversy since there is no agreement on its dependence on the sign of the temperature variation. The Uppsala group[4,5] claims that their measurements of the rate of change of the decay of the magnetisation are symmetric with respect to the sign of $\delta T$. On the other hand, the Saclay-UCLA group studies the decay of the magnetisation and claims that it has an explicit dependence on the sign of $\delta T$, *i.e.* that the response is asymmetric[6,7]. In Refs. 6 and 7 it is concluded that a short, as compared to the initial waiting time, heat pulse ($\delta T > 0$) partialy reinitializes aging while a negative temperature cycling ($\delta T < 0$) freezes the system into the state reached during the initial aging process. In other words, to lower the temperature is said to be equivalent to a, maybe partial, new quench. Instead, raising the temperature does not produce such an effect. (see Ref. 3 for a more detailed discussion on this discrepancy).

The out of equilibrium dynamics of the standard finite dimensional model for spin-glasses, *viz.* the 3-dimensional Edwards-Anderson model (3D-EA), has been extensively studied numerically[8–11]. Simulations for the TRM decay and for the two-time auto-correlation function decay at constant temperature[8,9] and after temperature changes during the waiting time[10,11] have been performed. In addition, simulations of the dynamics of the hypercubic spin-glass cell – a mean-field model for large dimensionalities – with a very schematic discussion of the effect of temperature shifts at $t_w$ have also been carried out[12].

Numerically, there is again agreement on the influence of the magnitude and duration of the heat pulses on the response of the 3D-EA, but there are different opinions as regards the symmetry or asymmetry in the response. In Ref. 10 it is claimed that the numerical results for the rate of change of the magnetisation[8] are symmetric. Conversely, in Ref. 11 numerical support for asymmetries in the response is given.

The aging effects and in particular the effects of temperature variations have been interpreted with various phenomenological models.

The Uppsala group interpreted their experimental results along the lines of droplet models[13–15], *i.e.* in terms of time-dependent domain growth with the added assumption of chaoticity. These models predict a symmetric response to temperature changes. The Saclay-UCLA group interpreted their asymmetric observations with a 'hierarchical' model[6] inspired by the replica solution of mean-field spin-glass models[16]. In this model it is assumed that the system relaxes in a rough energy landscape with an hierarchical organisation of many metastable states.

A semi–phenomenological approach to spin-glass dynamics based on the Parisi solution for the Sherrington–Kirkpatrick model has been proposed[17]. One constructs a Markov chain on a hierarchical tree with jump rates which are quenched random variables chosen from a Lévy distribution; the index of the distribution depending on the level to which the jump takes place. One associates an overlap between states depending on the ultrametric distance between them and then computes the correlation function within this framework. This model clearly exhibits aging phenomena and a simple two–level tree (corresponding to two step replica symmetry breaking) may be used to account for many of the experimental observations. Temperature changes are implemented by a reorganisation of the tree structure via changes in the Lévy law indices and overlap variables.

Only recently attention has been paid to the analysis of the off-equilibrium dynamics of mean-field spin-glass models. The long-time *analytic* solution of the dynamic equations of the $p$-spin spherical model[18], for $p \geq 3$, and for the Sherrington-Kirkpatrick model[19], have been worked-out. In addition a detailed numerical analysis of the dynamical equations for a particle moving in an infinite dimensional random potential with long-range correlations has been carried out[20] (see also Ref. 21). In these studies the dynamical equations are those associated to the relaxation of the system at constant temperature starting from a given initial condition. The effects of temperature variations have not yet been studied with mean-field spin-glass models.

In this paper we shall study the dynamics of the spherical Sherrington-Kirkpatrick model or the $p = 2$ spherical spin-glass model[22,23] for general initial conditions. The model is simple enough to be solved exactly statically and dynamically and allows us to examine in a simple analytical way some of the experimental scenarios.

From the static point of view the model is extremely simple. Its energy has only two minima, corresponding to the configurations with maximum (minimum) projection on the direction of the eigenvector associated to the maximum eigenvalue of the (Gaussian) interaction matrix. Within the replica formalism the model is solved exactly with a replica symmetric ansatz[22]; a result that would suggest its triviality from the dynamical point of view. Despite this, the model's dynamics is extremely interesting. For almost any initial condition it is out of equilibrium, and the evolution *does not* lead it to equilibrium. The model exhibits aging effects in the two-time auto-correlation function, *i.e.* even asymptotically it depends explicitly on the waiting time. Depending on the initial conditions, rather than the exponential types of decay in the energy-density, correlation functions, etc. one observes power-law decays. This is because the characteristic equilibration time of the system (for such initial conditions) is infinite. Such an exponential decay with rate given by the inverse of this equilibration time therefore cannot appear; the only other time–scale that can appear in correlation functions is the waiting time.

The aim of this paper is two-fold. On the one hand we probe some of the assumptions used to obtain analytical results for the long-time dynamics of more complicated mean-field spin-glass models[18,19]; namely, the weak-ergodicity breaking hypothesis[17]. On the other hand, we study the effects on this simple model of temperature jumps and temperature cyclings during the waiting time and we compare these analytical results with the associated experimental and numerical measurements. We do this via analysis of the decay of quantities such as the energy density and the auto-correlation function.

This being the 'simplest' mean-field spin-glass model, it is not expected to reproduce the experimental behaviour in every detail. Surprisingly enough, some of the strange features of real spin-glass behaviour are captured by it.

The paper is organized as follows. In Section II we present the model and describe the quantities of interest. In Section III we present the results for constant temperature $T(t) = T$. In Section IV we study the effects of temperature variations on the asymptotic decays. Finally, in Section V we present our conclusions.

## II. THE MODEL

The $p=2$ spherical model[22] is defined by the Hamiltonian

$$H = -\frac{1}{2}\sum_{i \neq j} J_{ij} s_i s_j \; , \tag{II.1}$$

where $s_i$, $i=1,\ldots,N$ are the spherical spin variables constrained such that $\sum_{i=1}^{N} s_i^2 = N$. The couplings between the spins are given by the quenched random variables $J_{ij}$.

To study the dynamical evolution it is convenient to diagonalize the coupling matrix $J_{ij}$ and to work with the time-dependent projections of the spin configuration $s(t) = \{s_i(t)\}$ onto the $J$-eigenvectors $\mu$: $s_\mu(t) = \mu \cdot s(t)$.

The dynamics for the model is defined via the usual Langevin equation, which when projected onto the eigenvectors $\mu$ becomes

$$\frac{\partial s_\mu(t)}{\partial t} = (\mu - z(t)) s_\mu(t) + h_\mu(t) + \xi_\mu(t) \; ; \tag{II.2}$$

$\mu$ is the eigenvalue associated to the eigenvector $\mu$, $h_\mu(t)$ represents an external magnetic field, $z(t)$ is a Lagrange multiplier enforcing the spherical constraint and $\xi_\mu(t)$ is the thermal noise with zero mean and correlation given by

$$\langle \xi_\mu(t) \xi_\nu(t') \rangle = 2\, T(t)\, \delta_{\mu\nu}\, \delta(t-t') \; . \tag{II.3}$$

$T(t)$ is the (possibly time-dependent) temperature. Hereafter we shall use $\langle \, \cdot \, \rangle$ to represent the average over the thermal noise.

The general solution to Eq. (II.2) is

$$s_\mu(t) = s_\mu(t_o)\, e^{\mu(t-t_o)}\, e^{-\int_{t_o}^{t} d\tau\, z(\tau)} + \int_{t_o}^{t} dt''\, e^{\mu(t-t'')}\, e^{-\int_{t''}^{t} d\tau'\, z(\tau')} \left( h_\mu(t'') + \xi_\mu(t'') \right) . \tag{II.4}$$

The two-time auto-correlation function is defined as usual: $C(t,t') \equiv (1/N) \left[ \sum_{i=1}^{N} \langle s_i(t) s_i(t') \rangle \right]_J$, where the square bracket indicates averaging over the disorder $J_{ij}$. It can be expressed in terms of the eigenvalues of $J$ as: $C(t,t') =$

$\int d\mu \; \rho(\mu) \; \langle s_\mu(t) s_\mu(t') \rangle$, with $\rho(\mu)$ the eigenvalue density. In the absence of an external field, using Eq.(II.4) one obtains

$$C(t,t') = \frac{1}{\sqrt{\Gamma(t)\Gamma(t')}} \left[ \langle\!\langle \, (s_\mu(t_o))^2 \, e^{\mu(t+t'-2t_o)} \, \rangle\!\rangle + 2 \int_{t_o}^{\min(t,t')} dt'' \, T(t'') \, \Gamma(t'') \, \langle\!\langle \, e^{\mu(t+t'-2t'')} \, \rangle\!\rangle \right]$$
(II.5)

where $\langle\!\langle \, \cdot \, \rangle\!\rangle$ stands for $\int d\mu \rho(\mu) \cdot$ and

$$\Gamma(t) \equiv \exp\left( 2 \int_{t_o}^{t} dt'' \, z(t'') \right)$$
(II.6)

may be computed self–consistently from the spherical constraint $C(t,t) = 1$, as the solution to the following Volterra equation of the second type:

$$\Gamma(t) = \langle\!\langle \, (s_\mu(t_o))^2 \, e^{2\mu(t-t_o)} \, \rangle\!\rangle + 2 \int_{t_o}^{t} dt'' \, T(t'') \, \Gamma(t'') \, \langle\!\langle \, e^{2\mu(t-t'')} \, \rangle\!\rangle \; .$$
(II.7)

This immediately implies $\Gamma(t_o) = 1$. In what follows we shall take the initial (quench) time to be zero: $t_o = 0$.

In this paper we shall analyse the case in which $J$ is a symmetric matrix with elements which are independently distributed Gaussian random variables with zero mean and variance proportional to $1/N$; this choice gives the model a well defined thermodynamic limit. The probability distribution function for the eigenvalues is then given by the Wigner semi–circle law[24]

$$\rho(\mu) = \frac{1}{2\pi} \sqrt{4 - \mu^2} \qquad \mu \in [-2, 2] \; .$$
(II.8)

However, as far as possible, we shall keep a general distribution $\rho(\mu)$. This will ultimately be useful in studying the effects of different, e.g. wide[25] distributions[26].

We study the long-time dynamics by calculating the long-time behaviour of

- The time dependence of the energy density and of its relaxation rate. In the absence of an external magnetic field one can show that the energy density is given by

$$\mathcal{E}(t) = \frac{1}{2}\left(T(t) - z(t)\right) = \frac{1}{2}\left(T(t) - \frac{1}{2}\frac{\partial}{\partial t}\ln(\Gamma(t))\right) \tag{II.9}$$

and its relaxation rate is given by

$$\frac{\partial \mathcal{E}(t)}{\partial \ln t} = \frac{t}{2}\left(\frac{\partial}{\partial t}T(t) - \frac{1}{2}\frac{\partial^2}{\partial t^2}\ln(\Gamma(t))\right) . \tag{II.10}$$

- The (averaged over the noise) staggered magnetisation

$$\langle s_\mu(t)\rangle = \frac{s_\mu(0)\, e^{\mu t}}{\sqrt{\Gamma(t)}} . \tag{II.11}$$

- The two-time correlation function (II.5). We shall be interested in the aging behaviour contained in $C(t, t')$ for $t$ and $t'$ large. Choosing $t \geq t'$, a useful expression for $C$ is the following:

$$C(t, t') = \frac{1}{\sqrt{\Gamma(t)\Gamma(t')}}\left[\Gamma(\frac{t+t'}{2}) - 2\int_{t'}^{\frac{t+t'}{2}} dt''\, T(t'')\, \Gamma(t'')\, \langle\!\langle e^{\mu(t+t'-2t'')}\rangle\!\rangle\right] . \tag{II.12}$$

We shall also study the decay from the initial conditions characterised by $C(t, 0)$.

- The equal-time overlap between two real replicas $s_\mu$ and $\sigma_\mu$ that evolve with the same thermal noise up to a 'waiting time' $t_w$, after which they evolve in two independent realisations of the thermal noise, say $\xi(t+t_w)$ and $\overline{\xi}(t+t_w)$:

$$Q(t+t_w, t+t_w) \equiv \langle\!\langle\, \langle s_\mu(t+t_w)\sigma_\mu(t+t_w)\rangle_{\xi,\overline{\xi}}\,\rangle\!\rangle$$

$$= \frac{1}{\sqrt{\Gamma_s(t+t_w)\Gamma_\sigma(t+t_w)}} \times$$

$$\langle\!\langle\, s_\mu(o)\sigma_\mu(o)\, e^{2\mu(t+t_w)} + 2\int_0^{t_w} dt''\, T(t'')\sqrt{\Gamma_s(t'')\Gamma_\sigma(t'')}\, e^{2\mu(t+t_w-t'')}\rangle\!\rangle$$
$$\tag{II.13}$$

The realisation of disorder is the same, *i.e.* the interaction matrix $J_{ij}$ is the same for both replicas.

- The response function at time $t$ to the perturbation by a small magnetic field applied at time $t'$

$$G(t,t') = \sum_{i=1}^{N} \frac{\delta \langle s_i(t) \rangle}{\delta h_i(t')}\bigg|_{\boldsymbol{h}=0}$$
$$= \langle\!\langle \sum_{\mu=1}^{N} \frac{\delta \langle s_\mu(t) \rangle}{\delta h_\mu(t')}\bigg|_{\boldsymbol{h}=0} \rangle\!\rangle \quad (\text{II.14})$$

$h_\mu(t) = \boldsymbol{\mu} \cdot \boldsymbol{h}(t)$. Taking the variation of $\langle s_\mu(t) \rangle$ with respect to $h_\mu(t')$ from Eq.(II.4) one obtains

$$G(t,t') = \sqrt{\frac{\Gamma_o(t')}{\Gamma_o(t)}} \langle\!\langle e^{\mu(t-t')} \rangle\!\rangle - \frac{1}{2} \frac{1}{(\Gamma_o(t))^{3/2}} \langle\!\langle \frac{\delta \Gamma_h(t)}{\delta h_\mu(t')}\bigg|_{\boldsymbol{h}=0} s_\mu(0) e^{\mu t} \rangle\!\rangle . \quad (\text{II.15})$$

The presence of a field modifies the function $\Gamma$. However, explicitly computing the functional derivative $\delta \Gamma_h(t)/\delta h_\mu(t')$ from the $\Gamma_h$ defining equation and taking the thermodynamic limit, one gets $\delta \Gamma_h(t)/\delta h_\mu(t')|_{\boldsymbol{h}=0} = 0$. Hence,

$$G(t,t') = \sqrt{\frac{\Gamma_o(t')}{\Gamma_o(t)}} \langle\!\langle e^{\mu(t-t')} \rangle\!\rangle , \quad (\text{II.16})$$

the subindex $o$ indicating the absence of an external field.

The decay of the TRM (assuming a suitably well behaved response function) is then given by

$$m_{t_w}(t) = h \int_0^{t_w} dt'' \, G(t,t'') . \quad (\text{II.17})$$

All these functions contain information about the nature of the dynamics and about the 'geometry' of the energy landscape of the model.

## III. CONSTANT TEMPERATURE

In this section we shall analyse the evolution of the system at constant temperature. For the initial conditions $s_i(0)$ at the quenching time $t_o = 0$ we choose two configurations that are representative of the different possible behaviours of the system:

- 'Uniform' initial condition.

$$s_\mu(0) = 1, \quad \forall \mu \quad \Rightarrow \quad \text{Non-equilibrium dynamics} \tag{III.1}$$

- 'Staggered' initial condition.

$$s_\mu(0) = \sqrt{\frac{\delta(a-\mu)}{\rho(a)}} \quad \Rightarrow \quad \begin{cases} \text{Non-equilibrium dynamics for } a \neq 2 \\ \text{Equilibrium dynamics for } \quad a = 2 \end{cases}$$

The 'uniform' initial condition that has a constant an equal to one projection onto each eigenvector of the interaction matrix $J$ is a random initial condition when written in the original basis. It corresponds then to the 'realistic experimental' initial condition from which the samples evolve after the rapid quench at $t_o$.

The strange form of the initial conditions in the staggered case follow from the requirement that the initial conditions satisfy the spherical constraint. In order to have an equilibrium dynamics, the initial condition must have a macroscopic condensation onto the maximum eigenvalue ($\mu = 2$ for $\rho(\mu)$ given by Eq.(II.8)). We shall show that only this particular initial condition leads the system to equilibrium, by studying the initial conditions needed to have a time-homogeneous relaxation of the correlation and response functions, two typical features of the equilibrium dynamics.

The function $\Gamma(t)$ can be obtained for general initial conditions using Laplace transform techniques to solve Eq. (II.7). We derive these results in Appendix A.

## A. Uniform initial condition

The function $\Gamma(t)$, for all times $t$ and $T < T_c = 1$, is given by

$$\Gamma(t) = \frac{1}{T} \sum_{k=0}^{\infty} k \, \frac{I_k(4t)}{2t} \, T^k \; . \tag{III.2}$$

For large times the energy density is

$$\mathcal{E}(t) \sim \frac{1}{2} T - 1 + \frac{3}{8 t} \tag{III.3}$$

and hence asymptotically it tends to the equilibrium value[22] $\mathcal{E}_{eq} = \frac{1}{2} T - 1$ with a power law decay.

The staggered magnetisation given by Eq. (II.11) behaves, asymptotically, as

$$\langle s_\mu(t) \rangle \sim (4\pi)^{1/4} \, q_{EA} \, (2t)^{3/4} e^{(\mu-2)t} \; , \tag{III.4}$$

$T < T_c$ and $q_{EA} = (1 - T)$ is the Edwards-Anderson order parameter[22]. If $\mu \neq 2$, it decays exponentially with time. Instead, if $\mu = 2$, the staggered magnetisation associated to the maximum eigenvalue grows with time as a power law: $\langle s_2(t) \rangle \sim t^{3/4}$. The system condenses, on average, onto the maximum eigenvalue.

The rate of decay from the initial conditions is characterised by

$$C(t,0) \sim \left(\frac{2}{\pi}\right)^{1/4} q_{EA} \, t^{-3/4} \; , \tag{III.5}$$

*i.e.* a power law decay of the correlations between the system and its initial conditions – this is typical of non–equilibrium dynamics.

The correlation function (II.5) is given by

$$C(t,t') = \frac{1}{\sqrt{\Gamma(t)\Gamma(t')}} \left[ \frac{I_1(2(t+t'))}{t+t'} + 2T \int_0^{t'} d\tau \, \Gamma(\tau) \frac{I_1(2(t+t'-2\tau))}{t+t'-2\tau} \right] \tag{III.6}$$

and asymptotically (large time $t'$) it behaves as

$$C(t,t') \sim 2\sqrt{2}\,\frac{\lambda^{3/4}}{(1+\lambda)^{3/2}}\,[1 - T\,\text{int}_1(2t(1-\lambda), 2t(1+\lambda))]\;, \tag{III.7}$$

where $\lambda = t'/t$ and

$$\text{int}_k(\alpha,\beta) \equiv \int_0^\alpha d\omega\,\frac{e^{-\omega}}{\left(1-\frac{\omega}{\beta}\right)^{3/2}}\frac{I_k(\omega)}{\omega}\;. \tag{III.8}$$

It is clear from Eq.(III.7) that even at zero temperature the correlation function shows an 'aging' behaviour: $C(t,t') = C(\lambda)$ and hence $C(t+t_w, t_w) = \tilde{C}(t/t_w)$. It can be also proven that $\partial C(t,t')/\partial t < 0$ and $\partial C(t,t')/\partial t' > 0$, for all times $t, t'$ such that $t > t'$. These are two of the 'weak-ergocicity' breaking properties[17,18].

For large $t$ there are three $t'$-regimes:

- Finite $t'$ or $\lambda \equiv t'/t \to 0$

$$C(t,t') \sim f(T,t')\,t^{-3/4} \tag{III.9}$$

with

$$f(T,t') = (4\pi)^{-1/4}\,q_{EA}\,2^{3/4}\,\frac{e^{2t'}}{\sqrt{\Gamma(t')}}\left[1 + 2T\int_0^{t'}dt''\,\Gamma(t'')\,e^{-4t''}\right]\;. \tag{III.10}$$

Having already taken the limit $t \to \infty$ one can now consider the limit $t' \to \infty$; thus

$$C(t,t') \sim 2\sqrt{2}\,q_{EA}\,\frac{\lambda^{3/4}}{(1+\lambda)^{3/2}} \sim 2\sqrt{2}\,q_{EA}\,\lambda^{3/4} \tag{III.11}$$

The correlation function decays to zero with a power law. Hence, given any finite time $t'$ there exists a big enough subsequent time $t$ such that the correlation function has decayed to zero.

- Large $t'$, $(t-t')/t \ll 1$, $\lambda = 1 - \left(\frac{t-t'}{t}\right) \to 1$; i.e. $t$ and $t'$ are relatively close to each other.

$$C(t,t') \sim C(t-t') = 1 - T\int_0^{2(t-t')}d\omega\,e^{-\omega}\,\frac{I_1(\omega)}{\omega}\;. \tag{III.12}$$

Up to zeroth order in $(t-t')/t$, the correlation function satisfies time-translation invariance (TTI). This is the so-called FDT (fluctuation dissipation theorem)-regime[18]. If $t = t'$, $\lambda = 1$ and $C(t,t) = 1$, as expected from the spherical constraint. If we take the limit $(t - t') \to \infty$ – determining the end of the FDT-scale – we obtain

$$C(t - t') \sim q_{EA} \qquad \text{(III.13)}$$

as can be clearly seen in Fig. 1.

This property completes the weak-ergodicity breaking scenario[17,19]. That is to say

$$\lim_{\tau \to \infty} C(\tau + t_w, t_w) = 0 \qquad \forall \ t_w \quad \text{finite}$$
$$\lim_{t_w \to \infty} C(\tau + t_w, t_w) = q_{EA} \qquad \forall \ \tau \quad \text{finite} .$$

The initial rate of decay of the correlation function is

$$\lim_{t' \to t} \frac{\partial C(t,t')}{\partial t} = -T . \qquad \text{(III.14)}$$

The usual fluctuation-dissipation theorem is satisfied for this range of times as we shall show below.

- Large $t'$, $(t-t')/t$ finite and $\lambda$ finite, *i.e.* widely separated times $t$ and $t'$.

  If $\lambda < 1$ the full Eq.(III.6) holds. For large $t$, $\text{int}_1(2t(1-\lambda), 2t(1+\lambda))$ is a function of $\lambda$. Hence, for $T < T_c = 1$ the correlation function depends explicitly on $t_w$. This is the 'aging' regime. The rates of decay $\partial C(t,t')/\partial t$ and $\partial C(t,t')/\partial t'$ are proportional to $-1/t$ and $1/t$, respectively, and hence the auto-correlation has a very slow variation.

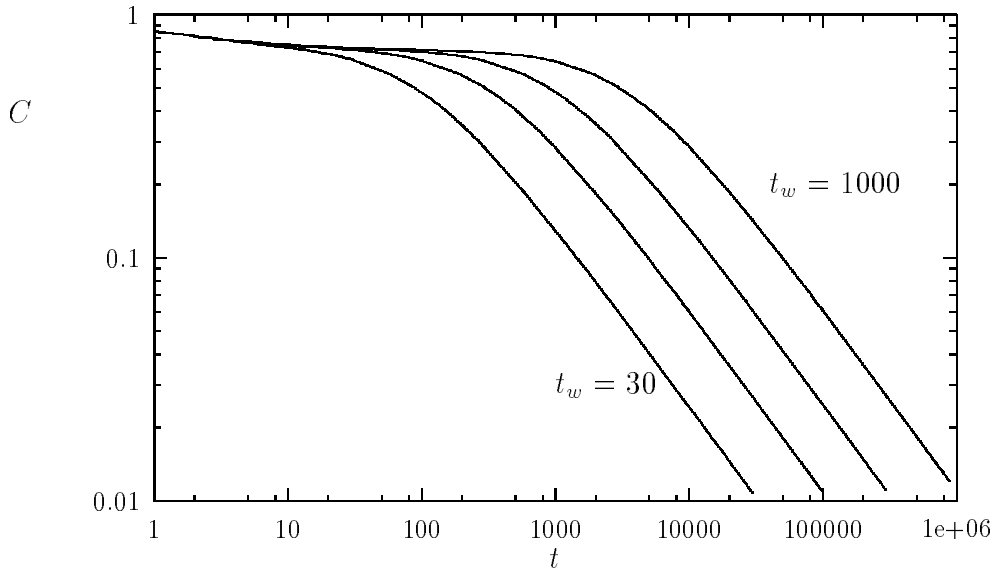

**Fig. 1** $C(\tau + t_w, t_w)$ vs. $\tau$ in a log-log scale. $t_w = 30, 100, 300, 1000$ (from left to right) at $T = 0.3$. $q_{EA}(T = 0.3) = .7$.

In Fig. 1 we present a plot of $C(\tau + t_w, t_w)$ vs. $\tau$ in a log-log scale, for constant temperature $T = 0.3$ and waiting times $t_w = 30, 100, 300, 1000, 3000$. The curves are the typical aging-curves already observed in the Monte Carlo simulations of the 3D Edwards-Anderson model[9] and the $D$-dimensional hypercubic spin-glass cell[12]. $\tau$ is the time elapsed after the 'waiting time' $t_w$. The waiting times in these figures are enough to show the asymptotic behaviour, the same behaviour is reproduced for longer waiting times. In all the curves two clearly different time regimes appear: for times $\tau$ much smaller than $t_w$ the auto-correlation has a fast decay from $C = 1$ to $C \sim q_{EA} = 1 - T = .7$, while for $\tau > t_w$ it has a slow decay from $C \sim q_{EA}$ to zero. The waiting time acts as the time-scale determining the length of the plateau in Fig. 1, *i.e.* the length of the FDT regime.

To compute the overlap between two (real) replicas we make them start from the same initial condition, the uniform one, and let them evolve in the way described in Section II. Then $\Gamma_s \equiv \Gamma_\sigma$ and

$$Q(\tau + t_w, \tau + t_w) = 1 - 2T \frac{1}{\Gamma(\tau + t_w)} \int_{t_w}^{\tau + t_w} dt'' \, \Gamma(t'') \frac{I_1(4(\tau + t_w - t''))}{2(\tau + t_w - t'')} \,. \qquad \text{(III.15)}$$

If $t_w = 0$ then for large times $\tau$, Eq. (III.15) implies:

$$Q(\tau, \tau) \sim q_{EA}^2 \ . \tag{III.16}$$

For a non-zero $t_w$ the behaviour depends on $t_w$. If $t_w$ is large one can show that there are two relevant $\tau$-regimes:

$$\text{If } 1 \ll \tau \ll t_w \quad Q(\tau + t_w, \tau + t_w) \sim \quad C(\tau + t_w, t_w)$$
$$\text{If } \tau \gg t_w \gg 1 \quad Q(\tau + t_w, \tau + t_w) \sim q_{EA} > C(\tau + t_w, t_w)$$

Fig. 2 shows, in a log-log scale, the $Q$ decay for $t_w = 0, 10, 30, 100$ at constant temperature $T = 0.6$. One can there see how the $\tau \to \infty$ limit of the curves depends on $t_w$, and how it approaches $q_{EA}$ for increasing $t_w$.

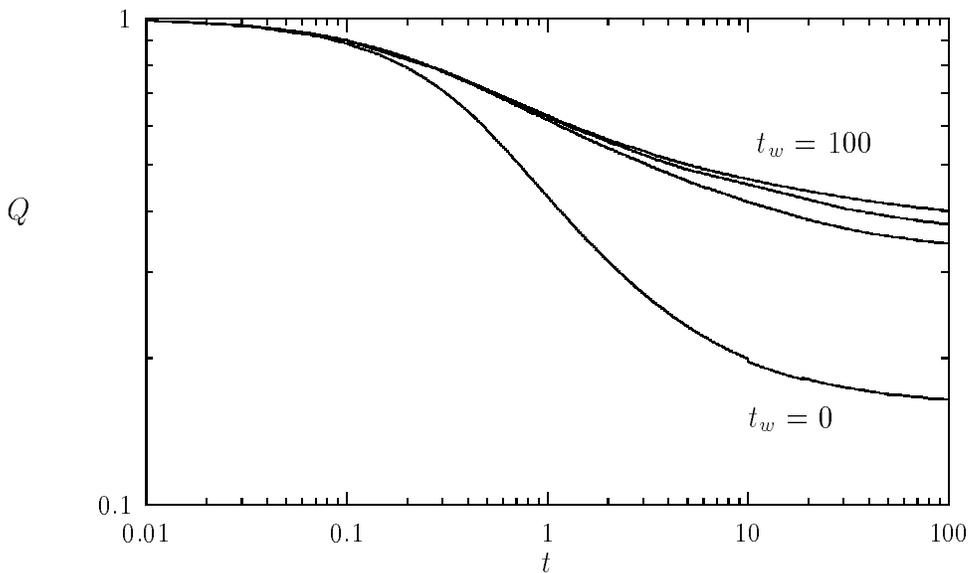

**Fig. 2.** The overlap $Q(\tau + t_w, \tau + t_w)$ vs. $\tau$ in a log-log scale at constant temperature $T = 0.6$ for $t_w = 0, 10, 30, 100$ (from bottom to top). $q_{EA}(T = 0.6) = 0.4$.

These results go in the same direction as those obtained numerically for the SK model[27] and analytically for the $O(N)$ model[26]. The systems escape from themselves faster than they do from each other. However, in this model, the asymptotic value of the overlap $Q$ at large $t_w$, when $\tau \to \infty$ is $q_{EA}$ while for the SK model[27], $Q(\tau + t_w, \tau + t_w) \sim 0$ when the same limits are considered. For the $p = 2$ spherical model this implies that there is a long–term drift in the energy landscape which causes the two replicas to follow roughly the same route. (One could imagine the replicas moving through channels or through the same sets of valleys in the energy landscape.) This demonstrates the relative simplicity of the energy landscape in the sense that a more chaotic landscape (*e.g.* that of the SK model) would possess a more complicated set of local–free energy minima which allow the replicas to become more widely separated asymptotically.

The response function reads

$$G(t, t') = \sqrt{\frac{\Gamma(t')}{\Gamma(t)}} \frac{I_1(2(t - t'))}{t - t'} \ . \tag{III.17}$$

For large times $t \geq t'$, the r.h.s. of (II.16) is independent of the temperature and

$$G(t, t') \sim t^{3/4} \, e^{-2t} \sqrt{\frac{I_1(4 t')}{t'}} \, \frac{I_1(2(t - t'))}{t - t'} \ . \tag{III.18}$$

One can demonstrate from this equation that:

- If $t'$ is finite the response function decays as the power law $G(t, t') \propto t^{-3/4}$.

- In the FDT scale $(t - t')/t \to 0$, $\lambda \to 1$,

$$G(t, t') \sim e^{-2(t-t')} \frac{I_1(2(t - t'))}{t - t'} \ , \tag{III.19}$$

*i.e.* it satisfies time-translation invariance. Furthermore, from Eqs.(III.12) and (III.19) one can immediately see that the FDT is also satisfied in this scale:

$$T \, G(t - t') = \frac{\partial C(t - t')}{\partial t'} \ . \tag{III.20}$$

- Finally, in the aging regime,

$$G(t, t') \sim \frac{\lambda^{-3/4}}{(1 - \lambda)^{3/2}} \, t^{-3/2} \ . \tag{III.21}$$

The response function decays as the power law $t^{-3/2}$. This makes a difference with the results obtained for the $p$-spherical model with $p \geq 3$, for which $G(t, t') \propto t^{-1}$. It implies that the memory of the $p=2$-model is too weak and the aging effects in the magnetisation are washed away quickly.

The TRM decay for large times $t$, if the magnetic field has been applied during a finite interval $[0, t_w]$, behaves as $m^{TRM}(t) \propto t^{-3/4}$ while if $t_w = \lambda_w t$, $\lambda_w > 0$, $m^{TRM}(t) \propto t^{-1/4}$.

One should note that many of the quantities calculated in this section depend on the initial conditions only via the term $\eta(\mu) = (s_\mu(0))^2$. For these quantities the results would be unchanged if one took any random initial conditions such that $\overline{\eta(\mu)} = 1$, the overline indicating the average over the randomness in the starting configuration (conditional on knowing the matrix $J$). For example the distributions $s_\mu(0) = \pm 1$ with equal probability and $s_\mu(0) \sim N(0,1)$ (i.e. zero mean Gaussian of unit variance) would have lead to the same results for such quantities. In addition the Gaussian initial conditions imply that $s_i(0) \sim N(0,1)$ by the rotational invariance of the Gaussian distribution.

Finally, let us remark that expressions (II.5) and (II.14) for the correlation and response functions solve, without the use of any assumptions, the dynamical mean-field equation of the $p$-spherical model when $p$ is set to 2.

In the language of Refs. 18 and 19 the long time dynamics is described in terms of two functions, namely, the function $X(C)$ that modifies the FDT

$$T\, G(t, t') = X(C(t, t')) \frac{\partial C(t, t')}{\partial t'} \tag{III.22}$$

and the 'triangular relation' $f$ that relates any three correlation functions at three long times

$$C(t, t') = f(C(t, t''), C(t'', t')), \tag{III.23}$$

$t > t'' > t'$. The fixed points of $f$, $f(a, a) = a$, separate different correlation scales.

For the $p = 2$ model we know the exact solution for all times and we can then compare the long-time part of it with the results obtained in Ref. 18 with the help of these assumptions. In fact, at large times, the exact solution implies the existence of two correlation scales.

In the first one the correlation decays from 1 to $q_{EA}$, $X(C) = 1$ and TTI and FDT are satisfied. This is the FDT scale. In the second one, the correlation decays from $q_{EA}$ to 0, $X(C) \sim t^{-1/2} \to 0$ and TTI and the FDT are violated. The result $X = 0$ agrees with the predicted value $X = (p-2)(1-q)/q$ of Ref. 18.

As regards the triangular relation the situation is slightly more subtle. The two correlation scales are separated by the value $C = q_{EA}$, hence, the three fixed points of $f$ are 0, $q_{EA}$ and 1. In general, just from its definition, between any two fixed points $f$ must have the form

$$C(t,t') = \jmath^{-1}\left(\jmath(C(t,t''))\,\jmath(C(t'',t'))\right), \tag{III.24}$$

with $\jmath$ a monotonous function to be determined by the model. There is going to be a function $\jmath$ for each correlation scale.

In the FDT scale the correlation functions are homogeneous functions of time and hence one can always write a relation like (III.24). One just uses the monotonicity of the correlation functions with respect to both times to invert the times in terms of the correlation functions: $C_{12} \equiv C(t_1, t_2) = C(t_1 - t_2) \Rightarrow t_1 - t_2 = C^{-1}(C_{12})$, then $C_{13} = C(C^{-1}(C_{12}) + C^{-1}(C_{23}))$ and finally $\jmath(C_{12}) = \exp(C^{-1}(C_{12}))$.

In the second scale $X = 0$ and the $p = 2$ long-time dynamic equations are identically satisfied without fixing the function $\jmath$. The most one can say about the correlation functions with the formalism used in Refs. 18 and 19 is that an equation like Eq.(III.24) exists.

It is interesting to note, however, that in the cases $p \geq 3$ the dynamical equations fix $\jmath$ to be the identity and then the triangular relation must be a product. They also imply $C(t,t') = q\, h(t')/h(t)$ with $h(t)$ a monotonous function of time. In the case $p = 2$, where we know the exact solution, this simple scaling clearly does not hold. Instead one has something more complicated which, e.g. for $T = 0$, can be written as

$$C(t,t') = \jmath^{-1}\left(\frac{h(t')}{h(t)}\right) \tag{III.25}$$

with $h(t) = t$ and $\jmath^{-1}(y) = 2\sqrt{2}\, y^{3/4}/(1+y)^{3/2}$. Only when $\lambda \sim 0$ one can write $C(\lambda) \sim q\, \lambda^{3/4}$.

## B. Staggered initial condition

In this section we shall consider the case where $T \neq 0$. When $T = 0$, if the system starts aligned in the direction of one of the eigenvalues of the interaction matrix then it is in a local energy minimum of the Hamiltonian and hence the dynamics is frozen. The computation of the functions $\Gamma$ for the various staggered initial conditions is presented in Appendix A.2.

In the case $a = 2$, the function $\Gamma$ leads to an exponential decay of the energy density towards $\mathcal{E}_{eq}$. When $a \neq 2$ the decay to $\mathcal{E}_{eq}$ is as in the case of uniform initial conditions.

The staggered magnetisation evolves in time as

$$\langle s_\mu(t) \rangle = \sqrt{\frac{\delta(\mu - a)}{\rho(a)}} \frac{1}{\sqrt{\Gamma(t)}} e^{\mu t} \qquad \text{(III.26)}$$

Then, at large times

$$a = 2 \quad \Rightarrow \quad \begin{cases} \langle s_{\mu \neq 2}(t) \rangle & \text{Stays zero } \forall t \\ \langle s_2(t) \rangle \to \sqrt{q_{EA}}\, s_2(0) \end{cases}$$

$$a \neq 2 \quad \Rightarrow \quad \begin{cases} \langle s_\mu(t) \rangle \sim q_{EA}\, s_a(0)\, e^{(\mu - 2)t}\, t^{3/4} \\ \langle s_{\mu \neq a}(t) \rangle & \text{Stays zero } \forall t \end{cases}$$

Thus, when the system starts from an initial configuration with a macroscopic condensation on the maximum eigenvalue, the averaged staggered magnetisation in this direction decays from its initial value to the asymptotic (macroscopic) one, with a weight $\sqrt{q_{EA}}$. Instead, when the system starts from different staggered initial conditions, all the averaged staggered magnetisations are zero, asymptotically.

The correlation with the initial conditions behaves as

- If $a = 2$

$$C(t, 0) \sim \sqrt{q_{EA}}. \qquad \text{(III.27)}$$

- If $a \neq 2$

$$C(t,0) \sim \frac{q_{EA}}{T^{\frac{1}{2}}} e^{-(2-a)t} \sqrt{2-a} \ (4\pi)^{\frac{1}{4}} (2t)^{\frac{3}{4}}. \qquad (\text{III.28})$$

Hence, in contrast from the case of uniform initial conditions, the decay of $C(t,0)$ is exponential if the system starts condensed on the eigenvalues $a \neq 2$, and $C(t,0)$ tends to a constant if $a = 2$.

In general for $\tau \gg t_w \gg 1$ one finds using Eq.(.16) that

- If $a = 2$

$$C(\tau + t_w, t_w) \sim q_{EA}. \qquad (\text{III.29})$$

- If $a \neq 2$

$$C(\tau + t_w, t_w) \sim q_{EA} \, 2\sqrt{2} \left( \frac{t_w(\tau + t_w)}{(\tau + 2t_w)^2} \right)^{\frac{3}{4}} = q_{EA} \, 2\sqrt{2} \left( \frac{2\lambda}{(1+\lambda)^2} \right)^{\frac{3}{4}}. \qquad (\text{III.30})$$

Hence for $a \neq 2$ the behaviour of the correlation function in this time regime exhibits explicitly the aging phenomenon. However, if $a = 2$ the system relaxes inside the equilibrium state of size $q_{EA}$. In addition one can show that the full form of the correlation function (for sufficiently large $t'$), starting from the staggered initial condition $a = 2$, is precisely the same as the correlation function for uniform initial conditions restricted to the FDT regime (see Eq(III.12)). Hence in all large time regimes it satisfies time translation invariance.

The two-replica overlap at $t_w = 0$, $Q(\tau, \tau)$, behaves as

$$\begin{aligned} a = 2 \quad & Q(\tau, \tau) \sim q_{EA} \\ a \neq 2 \quad & Q(\tau, \tau) \sim const \cdot q_{EA}^2 \, \frac{1}{T} \, e^{-2(2-a)\tau} \, \tau^{3/2} \end{aligned} \qquad (\text{III.31})$$

Hence, if the two replicas start from the 'equilibrium' initial condition, they evolve up to reaching a maximum distance of $q_{EA}$, the size of the equilibrium state. Instead if they start from any other staggered initial condition, they just separate completely, $Q(\tau, \tau)$ decays exponentially with time.

As far as the response function is concerned, it is easy to see that in the case where $a \neq 2$ the response behaves exactly as it does for the case of uniform initial conditions (for sufficiently large $t'$). For the case $a = 2$ one finds

$$G(t, t') \sim e^{-2(t-t')} \frac{I_1(2(t-t'))}{t - t'} \,, \tag{III.32}$$

hence it clearly exhibits time translational invariance and exponential decay for small time differences. As was the story for the correlation function, it has for all large $t'$, the same form as does the response function for uniform initial conditions restricted to the FDT regime. One may explicitly confirm that FDT is satisfied.

Thus, any staggered initial condition with $a \neq 2$ fails to reach an equilibrium regime. The staggered initial condition $a = 2$ is the only one leading to equilibrium dynamics – exponential decays, time-translation invariance.

Finally, let us mention that using the expression (.13) derived in Appendix A for the auto-correlation function one can show that a delta-type divergence in the initial condition at $\mu = 2$ is needed to ensure the equilibration of the system.

## IV. TEMPERATURE VARIATION EXPERIMENTS

In this Section we shall analyse the effects of temperature variations during the total waiting time in the asymptotic behaviour of the model. Since we are interested in non-equilibrium effects, namely aging effects, we shall let the system evolve from the uniform initial conditions of Section III, $\eta(\mu) = 1$, $\forall \mu$.

If the system starts at temperature $T$ at $t = 0$ and the temperature is changed to $T + \delta T$ at a later time $t_j$ then the function $\Gamma$ for this scenario is obtained as follows. For $t < t_j$, it is given by the constant temperature result, $\Gamma^{\text{jump}}(t) = \Gamma_T(t)$, as in Eq.(III.2). For $t > t_j$, the computation of Appendix B gives

$$\Gamma^{\text{jump}}(t) = \Gamma_{T+\delta T}(t) - 2\,\delta T \int_0^{t_j} dt'\, \Gamma_{T+\delta T}(t - t')\, \Gamma_T(t')$$
$$= \Gamma_T(t) + 2\,\delta T \int_{t_j}^{t} dt'\, \Gamma_{T+\delta T}(t - t')\, \Gamma_T(t') \,, \qquad (\text{IV.1})$$

In order to study this equation it is convenient to separate the 'asymptotic' factor $\exp(4t)$ (see Eq.(.7)) and to define the function $\gamma$:

$$\gamma(t) \equiv \exp(-4t)\, \Gamma(t) \,. \qquad (\text{IV.2})$$

For times $t$ such that $t - t_j \ll 1$

$$\Gamma^{\text{jump}}(t) \sim \Gamma_T(t)\left(1 + 2\,\delta T\,(t - t_j) + O((t - t_j)^2)\right) \,. \qquad (\text{IV.3})$$

For times $t$ such that $t \gg t_j \gg 1$ Eq.(IV.1) can be approximated

$$\gamma^{\text{jump}}(t) = \gamma_{T+\delta T}(t) - 2\,\delta T \int_0^{t_j} dt'\, \gamma_{T+\delta T}(t - t')\, \gamma_T(t')$$
$$\sim \gamma_{T+\delta T}(t)\left(1 - 2\delta T \int_0^{\infty} dt'\, \exp(-4t')\, \Gamma_T(t')\right)$$
$$= \gamma_{T+\delta T}(t)\left(1 - \frac{\delta T}{1 - T}\right) \,. \qquad (\text{IV.4})$$

For times $t_j \gg 1$ and $\delta t \equiv t - t_j \ll 1$:

$$\mathcal{E}^{\text{jump}}(t) \sim \mathcal{E}_{T+\delta T}(t) + \frac{1}{4}\left[\frac{\dot{\gamma}^2}{\gamma^2} - \frac{\ddot{\gamma}}{\gamma}\right]_{t_j}(t - t_j)$$
$$\sim \mathcal{E}_T(t) + (2 - T)\,\delta T\,\delta t\;. \qquad\text{(IV.5)}$$

It is therefore clear from above that the energy density is continuous at the temperature jump (as one would expect on physical grounds). At first order in $\delta T$ the effect of the temperature change is linear in $\delta T$. Fig. 3 shows the energy-density decay at constant temperatures $T = 0.6$ and $T = 0.9$. The dotted lines correspond to the equilibrium energy-densities $\mathcal{E}_T^{eq} = -0.7, -0.55$, respectively. In addition, the curve corresponding to the temperature cycle $T \to T + \delta T \to T$ with $T = 0.6$, $\delta T = 0.3$, $t_{w1} = 50$, $t_{w2} = 100$ is included. The effect of positive and negative temperature jumps can be seen in this curve. At $t_{w1}$ the first perturbation is applied and the energy density grows to a value above the asymptotic energy-density $\mathcal{E}_{T+\delta T}^{eq}$, in an interval such that $t - t_{w1}$ is short and then starts decaying towards the asymptotic value $\mathcal{E}_{T+\delta T}^{eq}$. A zoom in the figure would show that the perturbed curve for times bigger than $t_{w1}$ is above the asymptotic energy $\mathcal{E}_{T+\delta T}^{eq}$ and below the curve associated to the relaxation at constant temperature $T + \delta T$. It is an interesting feature of this system (and one that may be generic for non-equilibrium systems) that it reaches its final asymptotic energy density from above. Irrespective of the early time temperature pulse the time dependent energy density always becomes asymptotic to the energy density decay for a system that has been kept at constant (final) temperature. One can therefore say that a positive or negative temperature difference between the initial and final temperatures (irrespective of the duration of the pulse) has no effect on the ultimate rate at which the system reaches its final energy density.

For very large times $t$ as compared to $t_j$ the relaxation rate behaves as if the time-interval $[0, t_j]$ had not occurred:

$$\mathcal{E}^{\text{jump}}(t) \sim \mathcal{E}_{T+\delta T}(t) = \mathcal{E}_T(t) + \frac{1}{2}\,\delta T\;. \qquad\text{(IV.6)}$$

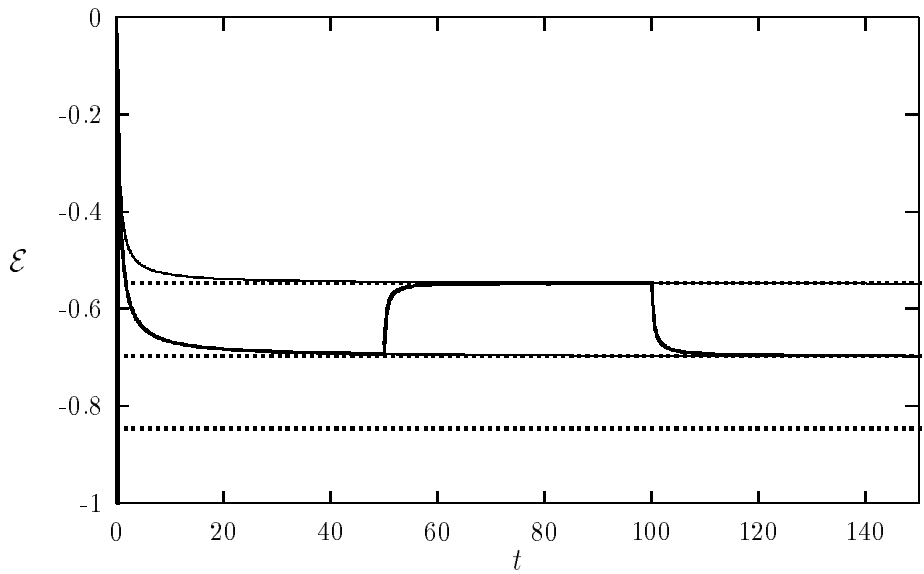

**Fig. 3.** The thiner curves correspond to the energy-density decays at constant temperature $T = 0.6, 0.9$. $\mathcal{E}_T^{eq} = -0.7, -0.55$, respectively. The dotted lines describe the two equilibrium energies. The thicker line corresponds to $\mathcal{E}(t)$ for the process $T \to T + \delta T \to T$, $T = 0.6$, $\delta T = 0.3$, $t_{w1} = 50$, $t_{w2} = 100$.

We now study the decay of the auto-correlation function when $t_j = t_w$. In this case, Eq. (II.5) implies

$$C^{\text{jump}}(t, t_w) = C_T(t, t_w) \sqrt{\frac{\Gamma_T(t)}{\Gamma^{\text{jump}}(t)}}, \qquad (\text{IV}.7)$$

where $C_T$ is the auto-correlation function at the 'initial' temperature $T$. The asymptotics for the auto-correlation function obtained in Section III therefore apply to the auto-correlation after a temperature jump, if one takes into account the presence of the multiplicative factor.

- Short times after the temperature jump - FDT scale.

  At the beginning of the FDT scale $t - t_w \ll 1$, $\Gamma^{\text{jump}}$ is given by Eq. (IV.3) and

  $$C^{\text{jump}}(t, t_w) \sim 1 - (T - 2\delta T)(t - t_w) + O((t - t_w)^2). \qquad (\text{IV}.8)$$

  Instead, if $t - t_w \to \infty$, one can show that

$$C^{\text{jump}}(t,t_w) \sim q_{EA}^{\text{jump}} \sim \sqrt{q_{EA}(T)\, q_{EA}(T+\delta T)} \qquad (\text{IV.9})$$

where $q_{EA}(T)$ is the Edwards-Anderson order parameter associated to temperature $T$.

- Large time-differences - Aging regime.

In the limit of large times $t \gg t_w$

$$C^{\text{jump}}(t,t_w) \sim C_T(t,t_w)\sqrt{\frac{q_{EA}(T+\delta T)}{q_{EA}(T)}} = C_T(t,t_w)\sqrt{1-\frac{\delta T}{1-T}}\,. \qquad (\text{IV.10})$$

It is clear from this expression that a positive (negative) temperature jump $\delta T > 0$ ($\delta T < 0$) implies a smaller (bigger) correlation function and then a younger (older) system than the one associated to the initial temperature. The $\delta T$-effect is linear at first order in $\delta T$.

These results can be seen explicitly in Fig. 4, where we present a plot of the auto-correlation function vs. $\tau$ ($\tau = t - t_w$), in a log-log scale, for $t_j = t_w = 300$ at constant temperature $T = 0.3, 0.6, 0.9$ and for the temperature jumps $T = 0.6$ and $\delta T = 0, \pm 0.1, \pm 0.3$.

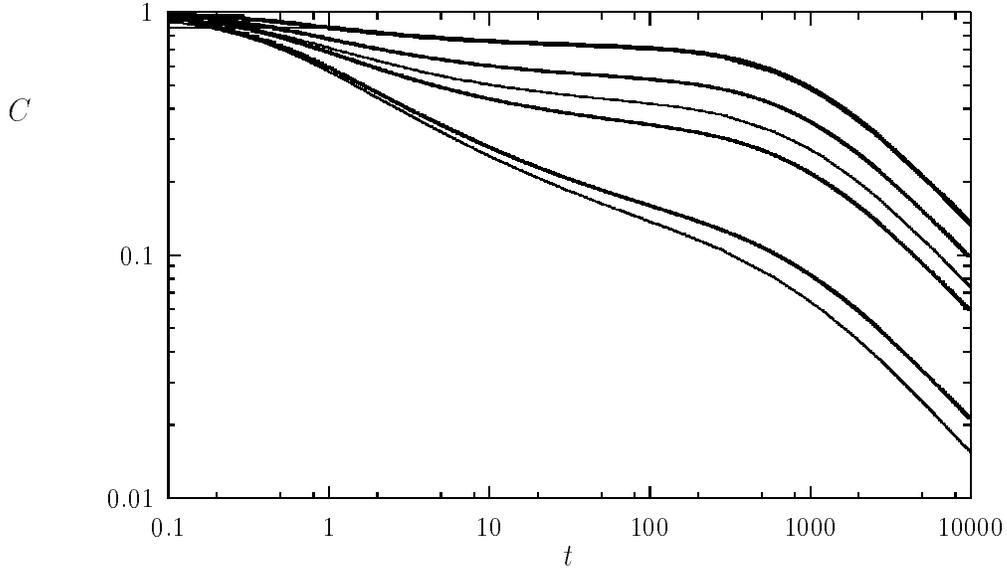

**Fig. 4.** Auto-correlation decay in positive and negative temperature jump experiments. Log-log plot of $C$ vs. $\tau = t - t_w$. $t_j = t_w = 300$ in all the curves. The thiner curves are associated to constant temperatures $T = 0.3, 0.6, 0.9$ from top to bottom. The thicker curves correspond to $T = 0.6$ and $\delta T = -0.3, -0.1, 0.1, 0.3$ from top to bottom, and hence to measuring temperatures $0.3, 0.5, 0.7, 0.9$ respectively.

More interesting is to compare the decay of the correlation function after the temperature jump with that associated to the final temperature $T + \delta T$, in the manner done in the experiments[4–6]. One sees that the curves after the jump get displaced in the direction of the older curves if $\delta T < 0$ or in the direction of younger curves if $\delta T > 0$. This result is equivalent to the displacement of the maximum of the logarithmic derivative of the magnetisation decay observed experimentally[4].

Nevertheless, this does not complete the understanding of the landscape of the model is concerned. The curves above are the analogous to the decay from the initial conditions $C(t,0)$ elsewhere in the paper but where the initial conditions are those generated by the evolution of the system at constant temperature $T$ for a time $t_w$.

If one performs the temperature jump at $t_j$ and then measures the $C(\tau + t_w, t_w)$ decay for $t_w \geq t_j$ then the rate of the correlation function decay can be said to depend on an effective age $t_a$ for the system at the final temperature. One would expect that $t_a \approx t_j \alpha + (t_w - t_j)$ for some positive value $\alpha$. In the case where $\alpha = 0$ then the effective age of the system is simply the time spent after the temperature change and hence the system has not benefitted from aging at the period $[0, t_j]$. For $t_w \gg t_j$ one may verify that $C^{\text{jump}}(t_w + \tau, t_w) \sim C_{T+\delta T}(t_w + \tau, t_w)$, i.e. the system has forgotten about the temperature jump. However if $\alpha$ is finite and $t_w \gg t_j$ then $C^{\text{jump}}(t_w + \tau, t_w) \sim C_{T+\delta T}(t_j \alpha + (t_w - t_j) + \tau, t_j \alpha + (t_w - t_j))$, hence this asymptotic analysis is not sufficiently sensitive to reveal the effective age induced by waiting at a different temperature. Instead one is forced to consider the intermediate range where $t_w \sim O(t_j)$. This computation has been carried out numerically and the results are shown in Figs 5.a and 5.b.

One sees that for $\delta T < 0$ (see Fig. 5.a.) the correlation function decay is very close to that of $C_{T+\delta T}(t_w + \tau, t_w)$, that is $\alpha \approx 1$ and hence the effective age of the system is close to $t_w$. In the case $\delta T > 0$ one sees that the correlation decay is actually slower than that for the system at fixed temperature, implying that the effective age of the system (viewed at the final measurement temperature) is in fact greater than $t_w$ and hence $\alpha > 1$.

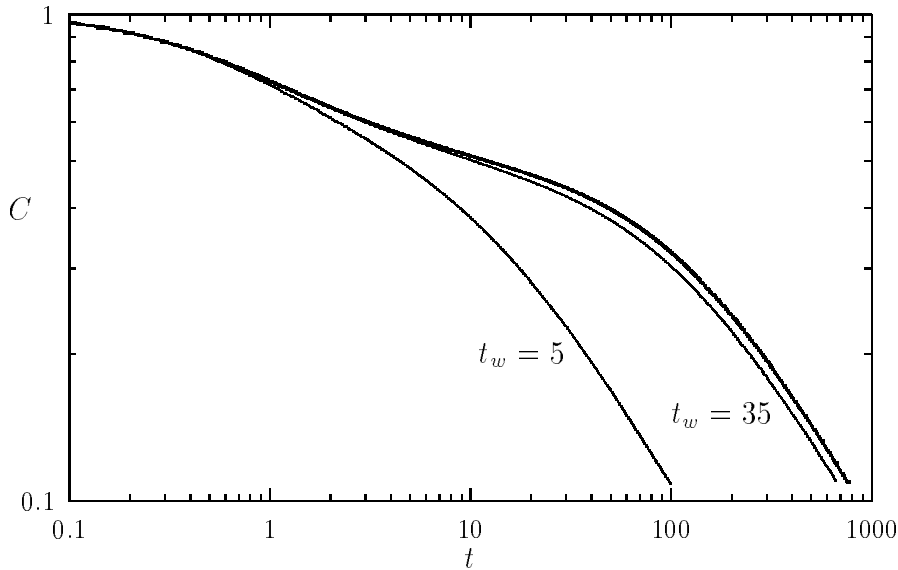

**Fig. 5.a** Negative temperature jump compared to constant temperature decays. $C(\tau + t_w, t_w)$ vs $\tau$ in a log-log scale. For the lower thin curve the temperature is constant, $T = 0.6$, and $t_w = 5$. For the upper thin curve the temperature is constant, $T = 0.6$, and $t_w = 35$, For the upper thick curve $t_j = 30$ and $t_w = 35$, $T = 0.9$, $\delta T = -0.3$, i.e. $T = 0.9 \to T + \delta T = 0.6$.

It has been observed experimentally in TRM and out-of-phase susceptibility decays[6] that a sample that has waited at a higher temperature ($\delta T < 0$) has an approximate effective age $t_a \sim t_w - t_j$, that is the period spent at the higher initial temperature does not contribute to the effective age of the system. Fig. 5.a. shows the correlation decay for such a setting. It is apparent from the curves that this model does not capture this feature. The lack of chaoticity in the energy landscape of the $p = 2$ spherical spin glass means that it fails to reproduce this important feature of real spin glasses.

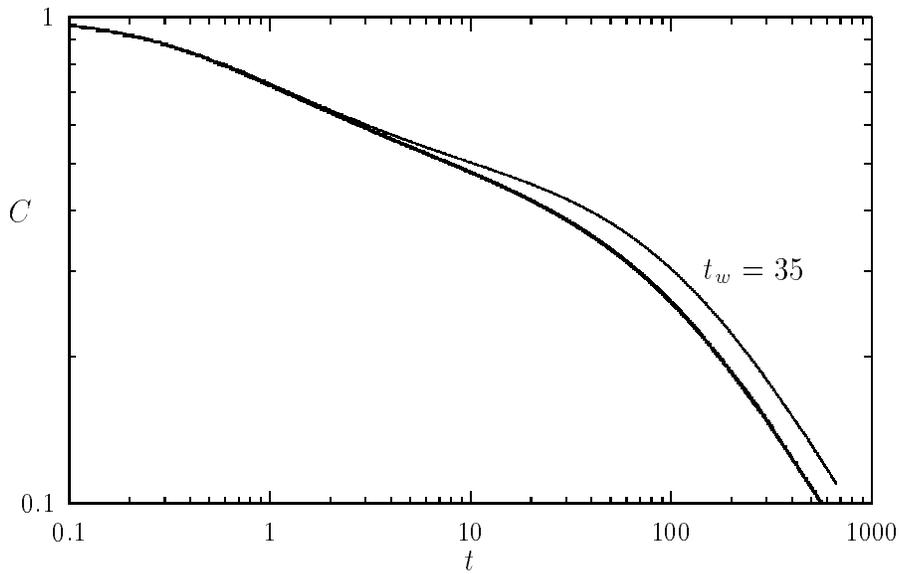

**Fig. 5.b**  Positive temperature jump. $C(\tau + t_w, t_w)$ vs $\tau$ in a log-log scale. For the thin continuous curve $t_w = 35$ and $T = 0.6$. For the thick curve $t_j = 30$ and $t_w = 35$, $T = 0.3$, $\delta T = 0.3$, i.e. $T = 0.3 \to T + \delta T = 0.6$.

The response of the model to temperature variation is represented by the results for temperature jumps. The model we are here considering is not able to describe some of the, though controversial, most interesting features of spin-glass physics as is the asymmetry in the response to positive or negative temperature variations.

One may also show that

$$Q^{\text{jump}}(\tau + t_w, \tau + t_w) = \frac{\Gamma_T(\tau + t_w)}{\Gamma^{\text{jump}}(\tau + t_w)} Q_T(\tau + t_w, \tau + t_w) \tag{IV.11}$$

if the temperature jump is applied at the same time $t_w$ at which the noise is changed. Thus, for large times $\tau \gg 1$

$$Q^{\text{jump}}(\tau + t_w, \tau + t_w) \sim \left(\frac{q_{EA}(T + \delta T)}{q_{EA}(T)}\right)^2 Q_T(\tau + t_w, \tau + t_w) . \tag{IV.12}$$

## V. CONCLUSIONS

In this paper we have examined the dynamics of the $p=2$-spherical model. In spite of having a very simple energy landscape, with only two equilibrium states and no signature of an exponential number of metastable states, its dynamical behaviour is far from being trivial.

For almost any initial condition the system never reaches equilibrium, though its energy density asymptotically approaches that of equilibrium. Some of the experimental observations are reproduced by the time-evolution of the auto-correlation function: a slow decay towards zero for widely separated times and aging effects represented by an explicit dependence on the waiting time.

However, up to now there are no experimental results about the auto-correlation decay; the measurements are concentrated on the magnetisation decay. For this simple model the magnetisation depends on $t_w$ but it decays too fast, with an extra factor $t^{-\alpha}$. It would we worth studying the effects of magnetic field variations in this solvable model[26].

A similar situation with aging effects in the correlation function and too fast a decay for the magnetisation has been obtained for 'unfrustrated' systems such as the XY model[28]. One may argue that the landscapes in which these models relax are far too simple and that this is the reason why the response function decays too fast. It is interesting to note that again, as it has already been noted in Ref. 28, the response appears to be more sensitive to the precise nature of the landscape than the auto-correlation function.

The analysis allowed us to check, in every detail, the validity for this model of the weak-ergodicity breaking and weak long term memory (rather too weak in this case) hypotheses.

The exact results demonstrate that, asymptotically, there are two time regimes. For close times compared to $t'$, $C$ decays from 1 to $q_{EA}$. TTI is satisfied both by $C$ and $G$ and the FDT theorem holds. For well separated times $C$ decays from $q_{EA}$ to 0, and TTI and the FDT theorem are violated. In the first regime $C$ decays quickly while in the second regime it decays slowly.

As regards the function $X[C]$ measuring the departure from the FDT theorem, one here finds $X[C] = 1$ if $q_{EA} \leq C \leq 1$ and $X[C] = 0$, if $0 \leq C < q_{EA}$ in good agreement with the extension of the results of Ref. 18 to the case $p = 2$.

In the case of initial conditions leading the system to equilibrium dynamics (*i.e.* with initial macroscopic condensation on the eigenvector associated to the maximum eigenvalue $\mu = 2$) we have been able to calculate the (time translationally invariant) autocorrelation and response functions. One finds that, somewhat surprisingly, these functions are *identical* to the corresponding functions in the FDT regime for initial conditions which ultimately lead to the aging phenomenon. It is not at all obvious that this should be the case and one is tempted to speculate on the possible generality of this phenomenon. It is possibly a consequence of a form of self similarity in the energy landscape; indeed we have also seen that all the initial conditions considered lead asymptotically to the same energy density[1], adding further weight to this notion.

We have also been able to examine the effects of temperature variations on the behaviour of the two-time autocorrelation function for this model. In real experimental situations one can assess a sort of effective age of a system that has experienced a temperature variation via TRM decay and a.c. susceptibility measurements [2] By analogy to the constant temperature

---

[1] This is not expected to happen when $p \geq 3$.

[2] To be more precise, in TRM experiments one can assign an effective age to a system that has undergone a negative cycling experiment. This is done by superposing the magnetisation decay curves after the cycle with a constant temperature decay for a smaller waiting time for all times explored experimentally; thus giving the effective waiting time. Nevertheless, this is not possible for positive cycling experiments, for which the decay at short times and at long times are considerably different, and the curves cannot be compared with constant temperature decays for any waiting

case, systems exhibiting a more rapid decay are said to be younger. A similar style analysis but using the correlation function to test the effective age has been carried out. Systems which have spent time at a higher temperature before switching to the final temperature *do* benefit from aging during time spent at higher temperature at variance with real spin glass experiments where the time spent at higher temperature does not contribute to the effective age of the system at the final temperature (if the temperature variation is big enough though all the temperatures are, of course, kept below the critical).

Preliminary investigations would suggest a similar behaviour for the $O(N)$ model. However, in this model one has a notion of space and we hope that a full study would allow us to develop an *analytic* picture of aging phenomena in terms of domain growth. Through this analysis we would like to attempt to make a connection between the droplet and the mean-field spin-glass models.

In one respect the $p = 2$ spin glass shares a property of droplet models. The existence of two well defined ground states (at least at zero temperature), parallel to the eigenvector with maximal eigenvalue, implies that the system evolves via a competition of these two phases. Indeed, in common with this model, droplet models are unable to account for the experimental effects of temperature cycling unless the additional hypothesis of chaoticity is added by hand.

It would be interesting to explicitly see how a complete 'aging' situation is settled when the parameter $p$, in the definition of the general $p$-spin spherical model, is raised from $p = 2$ to a bigger value[29,18]. The solution we have here presented for the $p = 2$ model, is the exact solution of the full mean-field dynamical equations of Ref. 18 when $p = 2$. Hence, it can be used as the unperturbed solution in a perturbative analysis of those equations around $p = 2$.

These final topics are currently under investigation[26].

---

time. See Ref. 3 for a more detailed discussion on this point.

ACKNOWLEDGEMENTS

We would like to thank G. Parisi for first drawing our attention to the interest of this model. During the course of this work we have benefited from useful discussions with J.-P. Bouchaud, J. Kurchan, M. Mézard and E. Vincent. L. F. C. wishes to thank the ITP at the University of California at Santa Barbara for hospitality.

# Appendix A

We here solve Eq.(II.7) at constant temperature $T$ for $\Gamma$; we shall assume that we are in the region $T < T_c$ ($T_c = 1$) – we shall *a posteriori* see that this guarantees the convergence of the expansions we use. The Laplace–transform of $\Gamma$ is given by

$$\tilde{\Gamma}(s) = \frac{\langle\langle \eta(\mu) \frac{1}{s-2\mu} \rangle\rangle}{1 - 2T \langle\langle \frac{1}{s-2\mu} \rangle\rangle}, \tag{.1}$$

where $\eta(\mu) \equiv (s_\mu(t_o))^2$. Taking the inverse–transform

$$\Gamma(t) = \sum_{k=0}^{\infty} d\mu\, \rho(\mu)\, \eta(\mu)\, e^{2\mu t} \int \prod_{i=1}^{k} d\lambda_i\, \rho(\lambda_i) \frac{1}{2(\mu - \lambda_i)}$$

$$+ \sum_{k=0}^{\infty} (2T)^k\, k \int d\lambda_1\, \rho(\lambda_1)\, e^{2\lambda_1 t} \int \prod_{i=2}^{k} d\lambda_i\, \rho(\lambda_i) \frac{1}{2(\lambda_1 - \lambda_i)} \int d\mu\, \rho(\mu)\, \eta(\mu) \frac{1}{2(\lambda_1 - \mu)} \tag{.2}$$

and re-summing the series

$$\Gamma(t) = \int d\mu\, \rho(\mu)\, \eta(\mu)\, e^{2\mu t} \frac{1}{1 - T\chi(\mu)}$$

$$+ 2T \int d\lambda\, \rho(\lambda)\, e^{2\lambda t} \int d\mu\, \rho(\mu)\, \eta(\mu) \frac{1}{2(\lambda - \mu)} \frac{1}{(1 - T\chi(\lambda))^2} \tag{.3}$$

where

$$\chi(\mu) \equiv PP \int d\lambda\, \frac{\rho(\lambda)}{\mu - \lambda}, \tag{.4}$$

and $PP$ indicates the Cauchy principal value. The initial condition $s_i(0)$ - or its 'staggered' distribution $\eta(\mu) = (s_\mu(0))^2$ - determines the time behaviour of $\Gamma$ and, in particular, its asymptotic behaviour.

If the interaction-matrix $J$ belongs to the Gaussian ensemble, the density of its eigenvalues is given by Eq.(II.8). For long times $t$ the integrals over $\mu$ are dominated by the maximum eigenvalue $\mu = 2$. Then, $\chi(2) = 1$ and

$$\Gamma(t) \sim \frac{e^{4t}}{\pi} \left\{ \frac{1}{q_{EA}} \int_0^4 d\epsilon\, \sqrt{\epsilon}\, \eta(2-\epsilon)\, e^{-2\epsilon t} + \frac{T}{q_{EA}^2} \int_0^4 d\epsilon\, \sqrt{\epsilon}\, e^{-2\epsilon t} \int d\mu\, \frac{\rho(\mu)\, \eta(\mu)}{2 - \epsilon - \mu} \right\} \tag{.5}$$

## A.1 $\mu$-Uniform Initial Conditions

If $\gamma_\mu(0) = 1$, $\forall \mu$, Eq.(.1) gives

$$\Gamma(t) = \frac{1}{T} \sum_{k=0}^{\infty} k \, \frac{I_k(4t)}{2t} \, T^k \,, \tag{.6}$$

for $T < T_c = 1$. $I_k$ are the generalized Bessel functions. Its asymptotic behaviour is

$$\Gamma(t) \sim \frac{1}{\sqrt{4\pi}} \frac{1}{(1-T)^2} \frac{e^{4t}}{(2t)^{3/2}} \,. \tag{.7}$$

## A.2 Staggered initial condition

If $\eta(\mu) = \frac{\delta(\mu-a)}{\rho(a)}$ then

$$\Gamma(t) = \sum_{k=0}^{\infty} (2T)^k \left\{ e^{2at} \left[ \int d\lambda \, \rho(\lambda) \frac{1}{2(a-\lambda)} \right]^k \right.$$
$$\left. + k \int d\lambda \, \rho(\lambda) \, e^{2\lambda t} \frac{1}{2(\lambda-a)} \left[ \int d\alpha \, \rho(\alpha) \frac{1}{2(\lambda-\alpha)} \right]^{k-1} \right\} \tag{.8}$$

and

$$\Gamma(t) = e^{2at} \frac{1}{1 - T\chi(a)} - T \int d\lambda \, \frac{\rho(\lambda)}{a-\lambda} \, e^{2\lambda t} \frac{1}{(1 - T\chi(\lambda))^2} \,. \tag{.9}$$

For long times the second term can be approximated. If $J_{ij}$ is in the Gaussian ensemble, the saddle-point method implies

$$\Gamma(t) \sim e^{2at} \frac{1}{1 - T\chi(a)} + \frac{T}{\pi} e^{4t} \int_0^4 d\epsilon \frac{\sqrt{\epsilon}}{2 - a + \epsilon} \, e^{-2\epsilon t} \frac{1}{(1 - T\chi(2-\epsilon))^2} \,. \tag{.10}$$

- If $a = 2$

$$\Gamma_{a=2}(t) \sim \frac{e^{4t}}{q_{EA}} \,. \tag{.11}$$

- If $a \neq 2$

$$\Gamma_{a \neq 2}(t) \sim \frac{T}{\sqrt{4\pi}} \frac{1}{q_{EA}^2} \frac{1}{2-a} \frac{e^{4t}}{(2t)^{3/2}} \ . \tag{.12}$$

### A.3 Computing the correlation function

Once we have solved for $\Gamma(t)$ we are left with the problem of computing the correlation function for non–zero waiting times $t_w$. In order to do this we must calculate the term

$$W(t, t_w) = \int_0^{t_w} d\tau \ \Gamma(\tau) \ \langle\langle \ e^{\mu(t+2t_w-2\tau)} \ \rangle\rangle \ . \tag{.13}$$

Taking the Laplace transform with respect to $t_w$ yields

$$\tilde{W}(t, s) = \tilde{\Gamma}(s) \ \langle\langle \frac{e^{\mu t}}{s - 2\mu} \rangle\rangle \tag{.14}$$

and using Eq. (.1) yields

$$\tilde{W}(t, s) = \frac{\langle\langle \frac{\eta(\mu)}{s-2\mu} \rangle\rangle \ \langle\langle \frac{e^{\mu t}}{s-2\mu} \rangle\rangle}{1 - 2T \ \langle\langle \frac{1}{s-2\mu} \rangle\rangle} \ . \tag{.15}$$

We now proceed as before, expanding the denominator then inverting the Laplace transform to obtain

$$W(t, t_w) = \frac{1}{2} \left[ \int d\mu d\mu' \ \frac{\rho(\mu)\rho(\mu') \ \eta(\mu) e^{2\mu t_w + \mu' t}}{(\mu - \mu')(1 - T\chi(\mu))} + \int d\mu d\mu' \ \frac{\rho(\mu)\rho(\mu') \ \eta(\mu) e^{2\mu' t_w + \mu' t}}{(\lambda - \mu)(\lambda - \mu')(1 - T\chi(\mu'))} \right.$$

$$\left. + \int d\mu d\mu' d\lambda \ \frac{\rho(\lambda)\rho(\mu)\rho(\mu') \ \eta(\mu) e^{2\lambda t_w + \mu' t}}{(\mu - \mu')(1 - T\chi(\lambda))^2} \right] \ , \tag{.16}$$

this form is convenient for obtaining the relevant asymptotics in the case $t \gg t_w$.

# Appendix B

In this Appendix we compute the function $\Gamma(t)$ for general cyclic variations of the temperature. We start by obtaining the result for the single temperature jump case: $T(t) = T\theta(t_1 - t) + (T + \delta T)\theta(t - t_1)$.

Let us define the function $f(t)$

$$f(t) \equiv \langle\langle e^{2\mu t} \rangle\rangle = \int d\mu\, \rho(\mu)\, e^{2\mu t} \tag{B.1}$$

Starting from a uniform initial condition Eq.(II.7) implies, for the single temperature jump experiment,

$$\Gamma^{\text{1-jump}}(t) = f(t) + 2T \int dt'\, f(t - t')\, \Gamma_{[0,t_1]}(t') + 2(T + \delta T) \int dt'\, f(t - t')\, \Gamma_{[t_1,\infty]}(t') \tag{B.2}$$

where

$$\Gamma_{[0,t_1]}(t') \equiv \Gamma(t)\theta(t_1 - t) \tag{B.3}$$

$$\Gamma_{[t_1,\infty]}(t') \equiv \Gamma(t)\theta(t - t_1)\,. \tag{B.4}$$

Solving the Laplace-transformed equation yields

$$\tilde{\Gamma}(s) = \frac{\tilde{f}(s)}{1 - 2(T + \delta T)\tilde{f}(s)} - \frac{2\delta T \tilde{f}(s) \tilde{\Gamma}_{[0,t_1]}(s)}{1 - 2(T + \delta T)\tilde{f}(s)} \tag{B.5}$$

and taking the inverse-transform gives

$$\Gamma^{\text{jump}}(t) = \Gamma_{T+\delta T}(t) - 2\delta T \int_0^{\min(t,t_1)} dt'\, \Gamma_T(t')\, \Gamma_{T+\delta T}(t - t')\,. \tag{B.6}$$

This solution should clearly yield $\Gamma^{\text{jump}}(t) = \Gamma_T(t)$ for $t < t_1$. This may be verified by noting that formally

$$\Gamma_T = (1 + 2Tf)^{-1} f\,, \tag{B.7}$$

where the products in the above expression are to be taken as the convolution of functions. Substituting this form for $\Gamma_{T+\delta T}$ into Eq. (B.6) yields the desired result.

For completeness sake we mention that with a similar argument one can obtain the function $\Gamma$ associated to $n$ temperature jumps

$$T(t) = \sum_{i=1}^{n} T_{i-1} \, \theta(t - t_{i-1}) \, \theta(t_i - t) + T_n \theta(t - t_n) \,. \tag{B.8}$$

It is given by

$$\Gamma^{n\text{-jump}} = \Gamma_{T_n}(t) - 2 \sum_{i=1}^{n} (T_n - T_{i-1}) \int_{t_{i-1}}^{t_i} dt' \, \Gamma_{T_n}(t - t') \, \Gamma^{(n-1)\text{-jump}}(t') \,. \tag{B.9}$$

In particular, if $n = 2$, $T_o = T_2 = T$, $T_1 = T + \delta T$ and $t > t_2$ we obtain the one cycle case:

$$\Gamma^{1\text{-cyc}}(t) = \Gamma_T(t) + 2\delta T \int_{t_1}^{t_2} dt' \, \Gamma_T(t - t') \, \Gamma^{1\text{-jump}}(t') \tag{B.10}$$

$$= \Gamma_T(t) + 2\,\delta T \int_{t_1}^{t_2} dt' \, \Gamma_T(t - t') \, \Gamma_{T+\delta T}(t')$$

$$- 4\,(\delta T)^2 \int_{t_1}^{t_2} dt' \, \Gamma_T(t - t') \int_0^{\min(t_1, t')} dt'' \, \Gamma_{T+\delta T}(t' - t'') \, \Gamma_T(t'') \,. \tag{B.11}$$